\documentclass[preprintnumbers,amsmath,amssymb,showpacs]{revtex4}

\usepackage{graphicx}
\usepackage{dcolumn}
\usepackage{bm}
\usepackage{graphicx}
\usepackage{booktabs}
\usepackage{CJK}
\usepackage{amsmath}
\usepackage{amssymb}
\usepackage{color} 

\begin{document}

\title{Bell-type inequality and quantum nonlocality in four-qubit  systems}
\author{Dong Ding$^{1,2}$}
\author{Fengli Yan$^1$ }
 \email{flyan@hebtu.edu.cn}
\author{ Ting Gao$^3$ }
 \email{gaoting@hebtu.edu.cn}
\affiliation {$^1$ College of Physics Science and Information Engineering, Hebei Normal University, Shijiazhuang 050024, China \\
$^2$Department of Basic Curriculum, North China Institute of Science and Technology, Beijing 101601, China \\
$^3$College of Mathematics and Information Science, Hebei Normal University, Shijiazhuang 050024, China}
\date{\today}

\begin{abstract}
We present a Bell-type inequality for four-qubit systems. Using the inequality we investigate quantum nonlocality of a generic family of states $\left|G_{abcd}\right\rangle$ [Phys. Rev. A \textbf{65}, 052112 (2002)] and several canonical  four-qubit entangled states.
It has been demonstrated that the inequality is maximally violated by so called  ``four-qubit maximal entangled state $\left|G_{m}\right\rangle$" and it is also violated by the four-qubit W state and a special  family  of states $\left|G_{ab00}\right\rangle$. Moreover, a useful entanglement-nonlocality relationship for the  family  of states $\left|G_{ab00}\right\rangle$ is obtained.

\end{abstract}

\pacs{03.65.Ud, 03.67.-a, 03.67.Mn}

\maketitle

\section{Introduction}

Entanglement is an inherent consequence in the multipartite quantum systems. As a physical resource, quantum entanglement has many important applications in quantum computation and quantum information \cite{QCQI2000}. It is worth noting that a fundamental question in the theory of entanglement is how to characterize and quantify entanglement \cite{HHHH2009, BYW2008, YGC2011, MCCSGH2011, HGY2012, GH2011,GH2012,GH2010}. In recent years, the characterization for multipartite entangled states has been considerably developed.
Since local unitary operations cannot affect the intrinsic nature of entanglement,  the nonlocal properties of a quantum state are related to entanglement.
The classification of a bipartite system under local unitary operators can be done by singular value decomposition \cite{QCQI2000}.
Under stochastic local operations and classical communication (SLOCC), a three-qubit entangled pure state can be converted either to the GHZ-state or to the W-state \cite{DVC2000, AACJLT2000}. In the case of four qubits entangled pure state, it was shown \cite{VDMV2002} that there exist nine classes of states.
Recently, for a pure multipartite entangled state, Liu \emph{et al} \cite{LLLQ2012} propose a method for the local unitary classification by using the symmetry  and tensor decomposition.
For quantification of the entanglement, as is well-known, one can take the concurrence \cite{concurrence1998} as a measure of bipartite entanglement states. In 2000, Coffman \emph{et al} \cite{Three-tangle2000} proposed the three-tangle which is a measure of genuine tripartite entanglement. Next the generalization of three-tangle to $n$-tangle for multi-qubit has been derived \cite{n-tangle-even2001, n-tangle-odd2009}. More recently, Sharma \emph{et al} reported \cite{Four-tangle2010, SS2013, Four-tangle2013} two expressions of four-tangle to quantify genuine four-body correlations by constructing polynomial invariants under local unitary operations.

Quantum theory allows correlations between spatially separated systems, while the local hidden variable (LHV) theory gives an upper bound to quantify nonlocality for a certain inequality.  A famous inequality as a way of verifying predications of quantum mechanics versus those of all possible LHV theories was first proposed by Bell \cite{Bell}. In 1969, Clauser \emph{et al} derived another correlation inequality, the Clauser-Horner-Shimony-Holt (CHSH) inequality \cite{CHSH1969}, which also provides a way of experimentally testing the LHV theory as an independent hypothesis separated from the quantum formalism. It has been proved that \cite{Gisin1991, GP1992} for the standard projective measurements all pure entangled states of two-qubit violate the correlation CHSH inequality. Thus the CHSH inequality can be used as a tool for testing the quantum formalism against the LHV theory and testing for entanglement within the quantum formalism. In the case of three parties Ac\'{\i}n \emph{et al} \cite{ACGKKOZ2004} considered a Bell inequality and showed that all three-partite pure entangled states violate it.
We noticed that there are two Bell-type inequalities for four qubits have been derived. One is SASA inequality \cite{SASA2005} that is maximally violated by the four-qubit cluster state and another is WYKO inequality \cite{WYKO2007} that is maximally violated by the four-qubit entangled state \cite{YC2006}
$\left|\chi\right\rangle=
     \frac{1}{2\sqrt{2}}(\left| {0000} \right\rangle  + \left| {1111} \right\rangle
         - \left| {0011} \right\rangle  + \left| {1100} \right\rangle
         - \left| {0101} \right\rangle  + \left| {1010} \right\rangle
         + \left| {0110} \right\rangle  + \left| {1001} \right\rangle).$
For the $N$-qubit case, the best-known inequality is the Mermin-Ardehali-Belinski-Klyshko (MABK) inequality \cite{Mermin1990, Ardehali1992, BK1993}, which is included in the Werner-Wolf-\.{Z}ukowski-Brukner (WWZB) inequality \cite{WW2001, ZB2002} as a special case and has been proved that quantum violation of this inequality increases with the number of particles.
Also, there are other inequalities used to characterize entanglement property, such as, Svetlichny inequality \cite{Svetlichny1987}, Mermin inequality \cite{Mermin1990, CGR2008}, Bell-type inequality \cite{AH2006, SU2007, CWKO2008, PE2009, LZCW2011, LF2012}, and so on.
Generally speaking, violation of these inequalities by entangled states is a signature of entanglement of quantum states and the magnitude of violation increases with the entanglement of state.

In this paper, we mainly study of the four-qubit entanglement and discuss the nonlocality for several canonical four-partite entangled states.
In the Hilbert space $(\mathbb{C}^2)^{\otimes 4}$, a general four-qubit pure state may be written as
\begin{equation}\label{}
    \left|\Psi^{ABCD}\right\rangle=\sum_{ijkl}a_{ijkl}|ijkl\rangle, ~~~~~~(i,j,k,l=0,1),
\end{equation}
where $a_{ijkl}$ are the complex coefficients with the basis vectors $|ijkl\rangle$, and $A, B, C$, and $D$ are the locations of qubits, respectively.
Consider only one class, the so called ``generic class" in nine classes for four-qubit entangled pure state \cite{VDMV2002}. More exactly, if  a generic pure state of four qubits can always be transformed to the generic family
 \begin{equation}\label{psi}
  \left|G_{abcd}\right\rangle=
     \frac{a+d}{2}(\left| {0000} \right\rangle  + \left| {1111} \right\rangle)
         + \frac{a-d}{2}(\left| {0011} \right\rangle  + \left| {1100} \right\rangle)
           +\frac{b+c}{2}(\left| {0101} \right\rangle  + \left| {1010} \right\rangle)
          + \frac{b-c}{2}(\left| {0110} \right\rangle  + \left| {1001} \right\rangle)
\end{equation}
under SLOCC, then  the pure state is in the generic class, where the complex parameters $a$, $b$, $c$, and $d$ satisfy the normalization condition $|a|^2+|b|^2+|c|^2+|d|^2=1$ and for the sake of  simplicity throughout this paper we only concern the cases where all $a$, $b$, $c$, and $d$ are real numbers.
We notice that although the four-tangle \cite{Four-tangle2010} of the state (\ref{psi}) is 1, i.e. $\tau_4=4|[(a_{0000}a_{1111}-a_{0111}a_{1000})+(a_{0011}a_{1100}-a_{0100}a_{1011})-(a_{0010}a_{1101}-a_{0101}a_{1010})-(a_{0001}a_{1110}-a_{0110}a_{1001})]^2|=1$, this class of the states contains four-qubit cluster-state, the GHZ-state, the two-EPR-pair-state, the maximal four-partite entangled state $|G_{m}\rangle$ (it will be defined later). Next we introduce a Bell-type inequality with different violations yielded by different entangled states. Especially, the maximal violation yields with the four-qubit maximal entangled state $|G_{m}\rangle$.

\section{Optimal Bell-type inequality for four-qubit maximal entangled state}

Based on the assumption of local realism, where all four qubits are locally but realistically correlated, we introduce a linear Bell-type inequality that is maximally violated by the four-qubit maximal entangled state
 \begin{eqnarray}\label{psi-m}
  \left|G_{m}\right\rangle & := & \left|G_{\frac{1}{\sqrt{2}}\frac{1}{\sqrt{2}}00}\right\rangle \nonumber \\
  & = & \frac{1}{2\sqrt{2}}(\left| {0000} \right\rangle  + \left| {1111} \right\rangle
         + \left| {0011} \right\rangle  + \left| {1100} \right\rangle
         +\left| {0101} \right\rangle  + \left| {1010} \right\rangle
         + \left| {0110} \right\rangle  + \left| {1001} \right\rangle).
\end{eqnarray}
In the structure of our inequality, we suppose that every observer is allowed to choose between two dichotomic observables. More precisely, there is an ensemble of four spatially separated qubits and the measurements $A_1=\mathbf{a}_1\cdot\hat{\sigma}$ or $A_2={\mathbf{a}}_2\cdot\hat{\sigma}$ are performed on qubit 1, $B_1={\mathbf{b}}_1\cdot\hat{\sigma}$ or $B_2={\mathbf{b}}_2\cdot\hat{\sigma}$ are performed on qubit 2, $C_1={\mathbf{c}}_1\cdot\hat{\sigma}$ or $C_2={\mathbf{c}}_2\cdot\hat{\sigma}$ are performed on qubit 3, and $D_1={\mathbf{d}}_1\cdot\hat{\sigma}$ or $D_2={\mathbf{d}}_2\cdot\hat{\sigma}$ are performed on qubit 4, where ${\mathbf{a}}_1, {\mathbf{a}}_2$, ${\mathbf{b}}_1, {\mathbf{b}}_2$, ${\mathbf{c}}_1, {\mathbf{c}}_2$, ${\mathbf{d}}_1$, and ${\mathbf{d}}_2$ are unit vectors, $\hat{\sigma}$ is   Pauli operator. We define the Bell-type operator
\begin{equation}\label{B}
\mathcal{B}:= A_1B_1C_1D_2-A_1B_2C_2D_2-A_2B_2C_1D_1-A_2B_1C_2D_1.
\end{equation}
Apparently, Eq.(\ref{B}) can be rewritten as $\mathcal{B}=A_1(B_1C_1-B_2C_2)D_2-A_2(B_2C_1+B_1C_2)D_1.$ In terms of absolute local realism, each outcome $X_i$ ($X=A,B,C,D$ and $i=1,2$) can either take value $+1$ or $-1$. It is easy to  obtain  the results that either $B_1C_1-B_2C_2=0$ and $B_2C_1+B_1C_2=\pm 2$, or $B_1C_1-B_2C_2= \pm2$ and $B_2C_1+B_1C_2=0$.

Suppose next that $p(a_1, a_2, b_1, b_2, c_1, c_2, d_1, d_2)$ is the probability that, before the measurements are performed, the system is in a state where $A_1=a_1$, $A_2=a_2$, $B_1=b_1$, $B_2=b_2$, $C_1=c_1$, $C_2=c_2$, $D_1=d_1$, and $D_2=d_2$.
Let correlation function $\langle \mathcal{B} \rangle$ denote the average over many runs of the experiment, then we have
\begin{eqnarray}\label{B_1}
|\langle \mathcal{B} \rangle_{\mathrm{LHV}}|
     & =   &|\sum_{a_1, a_2, b_1, b_2, c_1, c_2, d_1, d_2}p(a_1, a_2, b_1, b_2, c_1, c_2, d_1, d_2)(a_1b_1c_1d_2-a_1b_2c_2d_2-a_2b_2c_1d_1-a_2b_1c_2d_1)| \nonumber \\
     & \leq  &\sum_{a_1, a_2, b_1, b_2, c_1, c_2, d_1, d_2}p(a_1, a_2, b_1, b_2, c_1, c_2, d_1, d_2)|(a_1b_1c_1d_2-a_1b_2c_2d_2-a_2b_2c_1d_1-a_2b_1c_2d_1)| \nonumber \\
      & = & \sum_{a_1, a_2, b_1, b_2, c_1, c_2, d_1, d_2}p(a_1, a_2, b_1, b_2, c_1, c_2, d_1, d_2)\times 2   \nonumber \\
     & =    & 2.
\end{eqnarray}
Also,
\begin{eqnarray}\label{B_2}
|\langle \mathcal{B} \rangle_{\mathrm{LHV}}|
     & =   &|\langle A_1B_1C_1D_2-A_1B_2C_2D_2-A_2B_2C_1D_1-A_2B_1C_2D_1 \rangle|  \nonumber \\
     & =   &|\sum_{a_1, a_2, b_1, b_2, c_1, c_2, d_1, d_2}p(a_1, a_2, b_1, b_2, c_1, c_2, d_1, d_2)a_1b_1c_1d_2 \nonumber \\
     && -  \sum_{a_1, a_2, b_1, b_2, c_1, c_2, d_1, d_2}p(a_1, a_2, b_1, b_2, c_1, c_2, d_1, d_2)a_1b_2c_2d_2 \nonumber \\
     && -   \sum_{a_1, a_2, b_1, b_2, c_1, c_2, d_1, d_2}p(a_1, a_2, b_1, b_2, c_1, c_2, d_1, d_2)a_2b_2c_1d_1 \nonumber \\
     && -   \sum_{a_1, a_2, b_1, b_2, c_1, c_2, d_1, d_2}p(a_1, a_2, b_1, b_2, c_1, c_2, d_1, d_2)a_2b_1c_2d_1| \nonumber \\
     & =    & |\langle A_1B_1C_1D_2 \rangle-\langle A_1B_2C_2D_2 \rangle-\langle A_2B_2C_1D_1 \rangle-\langle A_2B_1C_2D_1 \rangle|.
\end{eqnarray}
Combining inequality (\ref{B_1}) and Eq. (\ref{B_2}) the following Bell-type inequality holds for the predetermined results:
\begin{equation}\label{inequality}
|\langle\mathcal{B}\rangle_{\mathrm{LHV}}|=|\langle A_1B_1C_1D_2 \rangle-\langle A_1B_2C_2D_2 \rangle-\langle A_2B_2C_1D_1 \rangle-\langle A_2B_1C_2D_1 \rangle| \leq 2.
\end{equation}
It means that  inequality (\ref{inequality}) holds for LHV theories.
\begin{figure}
  \centering\includegraphics[width=4in]{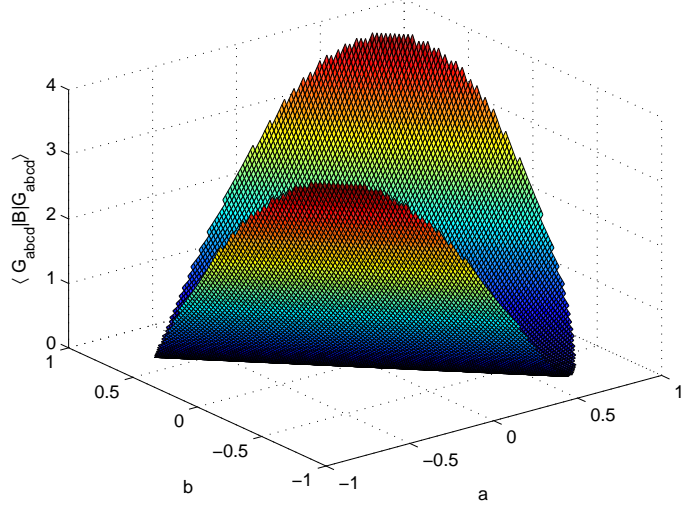}\\
  \caption{(color online). Plot of the quantum prediction $\langle G_{abcd}|\mathcal{B}|G_{abcd}\rangle$ versus $a$ and $b$ with $a^2+b^2\leq1$.}\label{P-B-ab}
\end{figure}

Then, we use the inequality to test the quantum nonlocality of the four-qubit maximally entangled state $|G_{m}\rangle$. Under the experimental setting
${\mathbf a}_1=\mathbf{b}_1=\mathbf{c}_1=\mathbf{d}_1=\mathbf{d}_2=(0,0,1)$, $\mathbf{a}_2=\mathbf{b}_2=\mathbf{c}_2=(0,1,0)$, the Bell-type operator reads
\begin{equation}\label{B-sigma}
\mathcal{B}=\sigma_z\sigma_z\sigma_z\sigma_z-\sigma_z\sigma_y\sigma_y\sigma_z-\sigma_y\sigma_y\sigma_z\sigma_z-\sigma_y\sigma_z\sigma_y\sigma_z.
\end{equation}
We notice that
\begin{equation}\label{}
\sigma_z\sigma_z\sigma_z\sigma_z|G_{m}\rangle=|G_{m}\rangle,
\end{equation}
and
\begin{eqnarray}\label{}
\sigma_z\sigma_y\sigma_y\sigma_z|G_{m}\rangle=-|G_{m}\rangle,\nonumber \\
\sigma_y\sigma_y\sigma_z\sigma_z|G_{m}\rangle=-|G_{m}\rangle,\nonumber \\
\sigma_y\sigma_z\sigma_y\sigma_z|G_{m}\rangle=-|G_{m}\rangle.
\end{eqnarray}
It is worth noting that there are 16 operators $\{S_i, i=1,\cdots, 16\}$  satisfying the eigenvalue equations $S_i\left|G_{m}\right\rangle=\pm\left|G_{m}\right\rangle$, namely, $\sigma_0\sigma_0\sigma_0\sigma_0, \sigma_x\sigma_x\sigma_x\sigma_x, \sigma_y\sigma_y\sigma_y\sigma_y, \sigma_z\sigma_z\sigma_z\sigma_z,
\sigma_x\sigma_x\sigma_0\sigma_0, \sigma_x\sigma_0\sigma_x\sigma_0, \sigma_0\sigma_x\sigma_x\sigma_0, \sigma_x\sigma_0\sigma_0\sigma_x, \sigma_0\sigma_x\sigma_0\sigma_x, \sigma_0\sigma_0\sigma_x\sigma_x,$
$\sigma_y\sigma_y\sigma_z\sigma_z, \sigma_y\sigma_z\sigma_y\sigma_z, \sigma_z\sigma_y\sigma_y\sigma_z, \sigma_y\sigma_z\sigma_z\sigma_y, \sigma_z\sigma_y\sigma_z\sigma_y,$ and $ \sigma_z\sigma_z\sigma_y\sigma_y$, where $\sigma_0$ is identity operator. From the above $16$ operators, we choose four operators and combine them as shown in Eq.(\ref{B-sigma}).
Quantum mechanically, therefore we derive the result of the expectation value of the Bell-type operator $\mathcal{B}$ stated in Eq.(\ref{B-sigma}) for the four-qubit maximally entangled state $\left|G_{m}\right\rangle$, that is,
\begin{equation}\label{}
\langle G_{m}|\mathcal{B}|G_{m} \rangle=4.
\end{equation}
This is indeed a violation of the inequality (\ref{inequality}) where the LHV bound is 2. Obviously, the maximum of $|\langle\mathcal{B}\rangle|$ is 4.  Under the optimal setting, the maximal violation of the Bell-inequality   reaches up to its maximum 4 for the four-qubit maximally entangled state $\left|G_{m} \right\rangle$. So our inequality, in a sense, is optimal and can also be acted as a strong entanglement witness for the entangled  state $\left|G_{m} \right\rangle$. This is also the reason why we call $\left|G_{m} \right\rangle$  the maximal entangled state.

Now, we study the quantum nonlocality of the generic family $\left|G_{abcd}\right\rangle$ using the above operator $\mathcal{B}$ stated by Eq.(\ref{B-sigma}). It turns out that
\begin{eqnarray}\label{B-ab}
\langle G_{abcd}|\mathcal{B}|G_{abcd}\rangle&=&\frac{1}{2}[(a+d)^2+(a-d)^2+(b+c)^2+(b-c)^2] +(b-c)(a+d)+(b+c)(a-d)  \nonumber \\
          &  &+(a-d)(a+d)+(b+c)(b-c)+(b+c)(a+d)+(b-c)(a-d)  \nonumber \\
          & =& 2(a+b)^2.
\end{eqnarray}
Under the constraint condition, $a^2+b^2\leq 1$, we plot the quantum prediction $\langle G_{abcd}|\mathcal{B}|G_{abcd}\rangle$ versus $a$ and $b$ as shown in Fig.\ref{P-B-ab}. Obviously, the violation of the inequality numerically depends on $a$ and $b$.

\section{Examples and discussion}

Next we  discuss several  examples of our Bell-type inequality for some special four-qubit entangled states.

(\uppercase\expandafter{\romannumeral1}) For the four-qubit maximal entangled state $\left|G_{m} \right\rangle$, the maximum violation reaches up to the maximum value $\langle G_{m}|\mathcal{B}|G_{m} \rangle=4$ and thus we claim that our inequality is optimal Bell-type inequality for the four-qubit maximal entangled state $\left|G_{m} \right\rangle$.

(\uppercase\expandafter{\romannumeral2}) For the four-qubit GHZ state $|\mathrm{GHZ}\rangle=|G_{\frac{1}{\sqrt{2}}00\frac{1}{\sqrt{2}}}\rangle=\frac{1}{\sqrt{2}}(\left| {0000} \right\rangle  + \left| {1111} \right\rangle)$, we obtain $\langle\mathrm{GHZ}|\mathcal{B}|\mathrm{GHZ}\rangle=1$. That is, the inequality is not violated by the four-qubit GHZ state.

(\uppercase\expandafter{\romannumeral3}) For the two-EPR-pair state, that is the state $\left|G_{1000}\right\rangle$, the quantum prediction is 2 and it does not violate the inequality.

(\uppercase\expandafter{\romannumeral4}) For the four-qubit cluster state $|\mathrm{cluster}\rangle=\frac{1}{2}(\left| {0000} \right\rangle +\left| {0011} \right\rangle+\left| {1100} \right\rangle - \left| {1111} \right\rangle)$, $\langle\mathrm{cluster}|\mathcal{B}|\mathrm{cluster}\rangle=1$, the inequality is not violated.

(\uppercase\expandafter{\romannumeral5}) For the entangled state $\left|\chi\right\rangle=
     \frac{1}{2\sqrt{2}}(\left| {0000} \right\rangle  + \left| {1111} \right\rangle
         - \left| {0011} \right\rangle  + \left| {1100} \right\rangle
         - \left| {0101} \right\rangle  + \left| {1010} \right\rangle
         + \left| {0110} \right\rangle  + \left| {1001} \right\rangle)$ investigated by Wu \emph{et al} \cite{WYKO2007}, $\langle\chi|\mathcal{B}|\chi\rangle=2$, it does not violate the inequality.

(\uppercase\expandafter{\romannumeral6}) For the state
$\left|\mathrm{HS}\right\rangle=
     \frac{1}{\sqrt{6}}[\left| {0011} \right\rangle  + \left| {1100} \right\rangle
         + \text{exp}({\text{i}\frac{2\pi}{3}})(\left| {0101} \right\rangle  + \left| {1010} \right\rangle)
         + \text{exp}({\text{i}\frac{4\pi}{3}})(\left| {0110} \right\rangle  + \left| {1001} \right\rangle)]$
introduced by Higuchi and Sudbery \cite{HS2000}, we have $\langle \mathrm{HS}|\mathcal{B}|\mathrm{HS}\rangle=0$ and the inequality is not violated.

(\uppercase\expandafter{\romannumeral7}) For the state
$\left|\Phi\right\rangle=
     \frac{1}{2}(\left| {0000} \right\rangle  + \left| {1101} \right\rangle)
         + \frac{1}{2\sqrt{2}}(\left| {0011} \right\rangle  + \left| {0110 } \right\rangle
         + \left| {1011} \right\rangle - \left| {1110} \right\rangle)$ as a candidate of maximally entangled state found by Brown \emph{et al} \cite{BSSB2005}, $\langle \Phi|\mathcal{B}|\Phi\rangle=\frac{1}{\sqrt{2}}$, the inequality is not violated.

(\uppercase\expandafter{\romannumeral8}) For the four-qubit W state $|\mathrm{W}\rangle=\frac{1}{2}(\left| {0001} \right\rangle +\left| {0010} \right\rangle+\left| {0100} \right\rangle + \left| {1000} \right\rangle)$, we obtain $|\langle\mathrm{W}|\mathcal{B}|\mathrm{W}\rangle|=2.5$, which means that the inequality is violated by the four-qubit W state.

\begin{figure}
  \centering\includegraphics[width=4in]{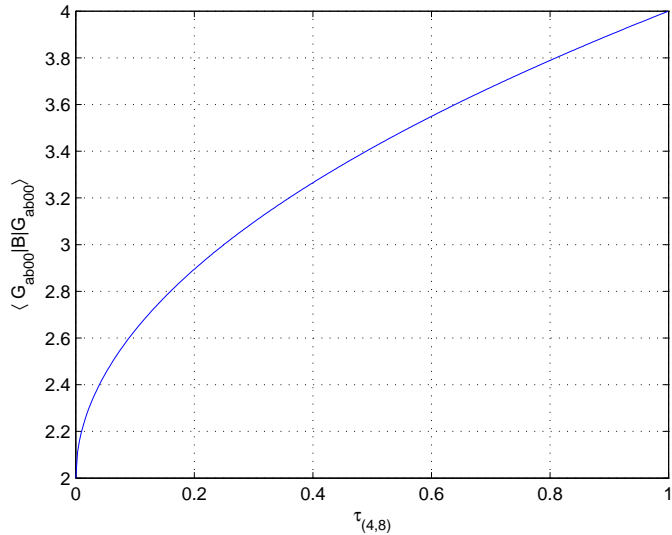}\\
  \caption{(color online). Numerical results of quantum nonlocality versus four-tangle for the family of the four-qubit entangled states $|G_{ab00} \rangle$, which violate inequality (\ref{inequality}).}\label{P-B-tau}
\end{figure}

A remarkable feature of our inequality is that  there exists a family of the four-qubit entangled states where all states violate the inequality. The family reads
 \begin{equation}\label{psi_ab}
  \left|G_{ab00}\right\rangle=
     \frac{a}{2}(\left| {0000} \right\rangle  + \left| {1111} \right\rangle
         + \left| {0011} \right\rangle  + \left| {1100} \right\rangle)
         +\frac{b}{2}(\left| {0101} \right\rangle  + \left| {1010} \right\rangle
         + \left| {0110} \right\rangle  + \left| {1001} \right\rangle),
\end{equation}
where $a$ and $b$ satisfy $ab>0$ and $a^2+b^2=1$. Consider genuine four-tangle for four-qubit states recently reported by Sharma \emph{et al} \cite{SS2013, Four-tangle2013}
\begin{eqnarray}\label{}
\tau_{(4,8)}(G_{abcd})& =& 4|\{[(a^2-b^2)(d^2-c^2)-(a^2-d^2)(b^2-c^2)]^2  \nonumber \\
          & &  +(a^2-b^2)(d^2-c^2)(a^2-d^2)(b^2-c^2)\}^{1/2}|,
\end{eqnarray}
where $\tau_{(4,8)}$ is genuine four-tangle and constructed by means of a four-qubit invariant of degree 8 expressed in terms of three-qubit invariants.
For the state $\left|G_{ab00}\right\rangle$, four-tangle is $\tau_{(4,8)}=4{a^{2}b^{2}}$. In the light of the result in Eq.(\ref{B-ab}) we obtain a useful relationship between four-partite entanglement (four-tangle) and nonlocality for the family of the four-qubit entangled states, that is,
\begin{equation}\label{}
\langle G_{ab00}|\mathcal{B}|G_{ab00}\rangle=2(1+\sqrt{\tau_{(4,8)}}).
\end{equation}
Fig.\ref{P-B-tau} shows the quantum nonlocality versus four-tangle for the family of the four-qubit entangled states $|G_{ab00}\rangle$.
It has been shown that quantum violation of the inequality (\ref{inequality}) varies with four-tangle $\tau_{(4,8)}$ and when $\tau_{(4,8)}=1$ (i.e., $a=b=1/\sqrt{2}$) the maximal violation yields.

\section{Summary}
In summary, we propose a four-qubit Bell-type inequality. The inequality is maximally violated by the four-qubit maximal entangled state $\left|G_{m}\right\rangle$ which do not violate SASA inequality \cite{SASA2005} and WYKO inequality \cite{WYKO2007}. We study the quantum nonlocality of the generic family $\left|G_{abcd}\right\rangle$ and several four-qubit states using our inequality and show that it is also violated by the four-qubit W state and a family of the four-qubit entangled states $\left|G_{ab00}\right\rangle$ while it is not violated by the four-qubit GHZ state, the four-qubit cluster state, the two-EPR-pair state and the state $\left|\chi\right\rangle$.
Also, we derive a useful relationship between four-partite entanglement and nonlocality for the family of the four-qubit entangled states $\left|G_{ab00}\right\rangle$. Note that our inequality is easier to test experimentally because it requires two settings and only four terms. Therefore, the specific derivation results in our inequality indicate that it can act as a strong entanglement witness for the family $\left|G_{ab00}\right\rangle$.
The discovery of the entanglement-nonlocality relationship, as well as the results reported in references \cite{GSDRS2009, AR2010, GDSKS2010}, will greatly facilitate the development of entanglement and their application in quantum information processing \cite{DLL2003, L-M-P2013, DY2013,  GaoTing, GaoYanWang, WY}.

\section*{Acknowledgements}
This work was supported by the National Natural Science Foundation of China under Grant No: 11371005,
Hebei Natural Science Foundation of China under Grant Nos: A2012205013.

\end{document}